# Thermally activated switching of perpendicular magnet by spin-orbit spin torque

Ki-Seung Lee[1,2], Seo-Won Lee[1], Byoung-Chul Min[2], and Kyung-Jin Lee[1,2,3a)]

[1]*Department of Materials Science and Engineering, Korea University, Seoul 136-701, Korea*

[2]*Spin Convergence Research Center, Korea Institute of Science and Technology, Seoul 136-791, Korea*

[3]*KU-KIST Graduate School of Converging Science and Technology, Korea University, Seoul 136-713, Korea*

**We theoretically investigate the threshold current for thermally activated switching of a perpendicular magnet by spin-orbit spin torque. Based on the Fokker-Planck equation, we obtain an analytic expression of the switching current, in agreement with numerical result. We find that thermal energy barrier exhibits a quasi-linear dependence on the current, resulting in an almost linear dependence of switching current on the log-scaled current pulse-width even below 10 ns. This is in stark contrast to standard spin torque switching, where thermal energy barrier has a quadratic dependence on the current and the switching current rapidly increases at short pulses. Our results will serve as a guideline to design and interpret switching experiments based on spin-orbit spin torque.**

---

[a)] Electronic mail: kj_lee@korea.ac.kr.

Spin transfer torque (STT) [1, 2], caused by spin-angular-momentum transfer from conduction electron spin to local magnetization, offers a way to reverse the magnetization direction by means of an electric current. In spin-valve structures consisting of two ferromagnetic layers (i.e., a switchable free layer and a fixed layer) separated by a normal metal spacer or an insulating layer, STT can change the stack resistance through current-induced magnetization switching of the free layer [3-7], which is used as the key working principle of magnetic random access memories (MRAMs).

For ultrahigh density MRAMs, a perpendicularly magnetized free layer is superior to an in-plane layer as it allows a smaller switching current for a given thermal stability factor [8-10]. Current-induced magnetization switching of a perpendicular layer can be achieved by one of the two ways. A conventional way is to inject a current perpendicular to magnetic tunnel junctions containing a perpendicular fixed layer and a tunnel barrier, which we call below "standard STT switching". An alternative way is to inject an in-plane current through a bilayer structure composed of a perpendicular free layer and a heavy metal layer [11, 12]. This in-plane current induced switching is realized by the so-called spin-orbit spin transfer torque (SO-STT). The dominant mechanism of the SO-STT is under intense discussion; it results from either the bulk spin Hall effect [13-15], or the interfacial Rashba-type spin-orbit coupling effect [16-20], or both [21-24].

From the application point of view, this SO-STT switching has the pros and cons compared to the standard STT switching. The most attractive feature of the SO-STT switching is that a large current density does not pass through the tunnel barrier, since it is operated by an in-plane current. This feature allows increasing the tunnel barrier thickness, which in turn increases the tunnel magnetoresistance of MRAM cells and the breakdown voltage of the tunnel barrier. The read, write, and breakdown voltages in MRAM operations always have cell-to-cell variations in the wafer level and distributions even in repeated switching of a single MRAM cell; typically |read voltage| < |write voltage| < |breakdown voltage|. In the standard STT switching, the interference among the read, write, and breakdown voltages causes MRAMs to malfunction so that the overlapping of these three voltage distributions should be minimized. By contrast, in the SO-STT switching, only the interference between the read and breakdown voltages matters, since the writing is governed mainly by an in-plane current. Another important merit of the SO-STT switching is that it allows a faster switching than the standard STT switching since the SO-STT switching is completed through only half a precession in contrast to the standard STT switching [25].

On the other hand, several challenging issues of the SO-STT switching still remain to be overcome. For instance, the SO-STT switching requires a rather large switching current density [11, 12]. This issue may be resolved by enhancing the effective spin Hall angle [26] and/or accompanying to the voltage-induced anisotropy change [25, 27-30]. A possible architecture of perpendicular SO-STT MRAM consists of multiple MRAM cells sharing the same heavy metal line and select-transistors (or diodes) to apply a voltage to individual cells (see Fig. 1). A selected MRAM cell on the heavy metal line can be written with a reasonably small in-plane current by turning on the voltage bias of the selected cell to induce the voltage-induced anisotropy reduction [25]. This SO-STT switching scheme however raises another issue that all MRAM cells sharing the same heavy metal line are half-selected, leading to an additional constraint in terms of write error; this half-selection issue has been serious in the field-driven switching scheme [31].

Because of the outstanding advantages and possibly manageable disadvantages mentioned above, the SO-STT switching recently attracts considerable interest from spintronics community, leading to detailed studies about it [32-39]. For the device application, the analytic expression of the switching current $I_\text{C}^\text{SO-STT}$ is essential, which was reported by us as [25]

$$I_\text{C}^\text{SO-STT} = \frac{e}{\hbar}\frac{M_S t_\text{F} H_\text{K,eff} A_\text{HM}}{\theta_\text{SH}^\text{eff}}\left(\sqrt{\frac{1}{8}\left[8 + 20 h_x^2 - h_x^4 - h_x\left(8 + h_x^2\right)^{3/2}\right]}\right), \quad (1)$$

where $M_\text{S}$ is the saturation magnetization, $t_\text{F}$ is the thickness of free layer, $\theta_\text{SH}^\text{eff}$ is an effective spin Hall angle, $A_\text{HM}$ is the cross-sectional area of heavy metal layer, $H_\text{K,eff}$ is the effective perpendicular anisotropy field, $h_x = H_x/H_\text{K,eff}$, and $H_x$ is the in-plane magnetic field required for the selective switching [11, 12]. Note that Eq. (1) was derived at zero temperature. Since experiments are always performed at a finite temperature, an important question is how thermal fluctuations affect the threshold current of the SO-STT switching. For example, the relationship between the thermal energy barrier and SO-STT is key information to design MRAM cells free from unwanted switching of half-selected cells at ambient temperature. The influence of SO-STT on the thermal energy barrier is also critically important to estimate the effective spin Hall angle or the magnitude of current-induced effective field from switching experiments done at a finite temperature.

In this work, we theoretically study thermally activated switching induced by SO-STT based on the Fokker-Planck equation. We focus on the single domain switching assuming that the lateral size of MRAM cell is smaller than the domain wall width, which is required for ultrahigh density MRAM beyond 1 Gbit. We provide a complete set of equations for the thermal energy barrier modified by SO-STT. We also provide an approximated analytic form of the thermal energy barrier, which is easy to be implemented in designing and interpreting switching experiments.

Magnetization switching induced by SO-STT is described by the Landau-Lifshitz-Gilbert equation with including spin-orbit damping-like spin torque term as [11, 14]

$$\frac{\partial \mathbf{m}}{\partial t} = -\gamma \mathbf{m} \times \mathbf{H}_{\text{eff}} + \alpha \mathbf{m} \times \frac{\partial \mathbf{m}}{\partial t} + \gamma c_J \mathbf{m} \times (\mathbf{m} \times \hat{\mathbf{y}}), \qquad (2)$$

where $\mathbf{m} = (\sin\theta\cos\varphi, \sin\theta\sin\varphi, \cos\theta)$ is the unit vector along the free-layer magnetization, $\gamma$ is the gyromagnetic ratio, $\alpha$ is the Gilbert damping constant, $c_J$ is $(\hbar/2e)(\theta_{\text{SH}}^{\text{eff}} J / M_S t_F)$, and $J$ is the current density. We neglect spin-orbit field-like spin torque for simplicity. $\mathbf{H}_{\text{eff}} = -\partial E / \partial (M_S \mathbf{m})$ is the effective magnetic field and $E$ is the magnetic energy density given as

$$E = -H_{\text{K,eff}} M_S \left( h_x \sin\theta\cos\varphi + \frac{1}{2}\cos^2\theta \right). \qquad (3)$$

The Fokker-Planck equation is given by [40]

$$\frac{\partial W}{\partial t} = -\frac{\partial}{\partial q}\frac{dq}{dt}W + \frac{D}{1+\alpha^2}\frac{\partial}{\partial q}g^{-1}\frac{\partial}{\partial q}W - \frac{\partial}{\partial p}\frac{dp}{dt}W + \frac{D}{1+\alpha^2}\left(\frac{M_S}{\gamma}\right)^2 \frac{\partial}{\partial p}g\frac{\partial}{\partial q}W, \qquad (4)$$

where $W$ is the distribution function of the magnetization in the $(q, p)$ phase space, $q$ is the canonical coordinate, $p$ is the momentum coordinate, $g = \sin^2\theta$, $D = \alpha\gamma k_B T /(M_S V)$ is the diffusion coefficient, $T$ is the temperature, and $V$ is the free-layer volume. Following Refs. [41] and [42], we obtain the Lagrangian density $L$ from Eq. (4) as

$$L = -\lambda_E \frac{dE}{dt} - H_E, \qquad (5)$$

where $H_E$ is the Hamiltonian density given as

$$\begin{aligned}
H_E = {} & \lambda_E \frac{\partial E}{\partial q}\left[\frac{\alpha M_S c_J}{1+\alpha^2}\frac{\partial}{\partial p}\mathbf{m}\cdot\hat{\mathbf{y}} + \frac{\alpha\gamma}{(1+\alpha^2)M_S}g^{-1}\frac{\partial E}{\partial q} + \frac{\gamma c_J}{(1+\alpha^2)}g^{-1}\frac{\partial}{\partial q}\mathbf{m}\cdot\hat{\mathbf{y}}\right] \\
& -\lambda_E \frac{\partial E}{\partial p}\left[\frac{\alpha M_S c_J}{1+\alpha^2}\frac{\partial}{\partial q}\mathbf{m}\cdot\hat{\mathbf{y}} - \frac{\alpha M_S}{(1+\alpha^2)\gamma}g\frac{\partial E}{\partial p} - \frac{M_S^2 c_J}{(1+\alpha^2)\gamma}g\frac{\partial}{\partial p}\mathbf{m}\cdot\hat{\mathbf{y}}\right] \\
& + \lambda_E^2 \frac{\alpha}{1+\alpha^2}g^{-1}\left(\frac{\partial E}{\partial q}\right)^2 + \lambda_E^2 \frac{\alpha}{1+\alpha^2}g\left(\frac{M_S}{\gamma}\right)^2\left(\frac{\partial E}{\partial p}\right)^2,
\end{aligned} \qquad (6)$$

and $\lambda_E$ is a variable describing the switching path. Using $q = \varphi$, $p = \partial \dot{L}/\partial \dot{q} = M_S(1-\cos\theta)/\gamma$, and $\varphi = 0$ for the switching path of SO-STT switching [25], we obtain

$$H_E = -\lambda_E \gamma \frac{\alpha}{(1+\alpha^2)} M_S H_{K,\text{eff}} (-h_x \cos\theta + \cos\theta \sin\theta)(h_S + h_x \cos\theta - \cos\theta \sin\theta)$$
$$+ \lambda_E^2 \frac{\alpha}{1+\alpha^2} M_S^2 H_{K,\text{eff}} (-h_x \cos\theta + \cos\theta \sin\theta)^2, \quad (7)$$

where $h_S = c_J / H_{K,\text{eff}}$.

In the low temperature limit, the thermal energy barrier is obtained by integrating the Lagrangian density along the optimal path [41, 42]. By solving $H_E = 0$, one finds the optimal path $\lambda_E^*$, that is, the solution of $\lambda_E$ satisfying $H_E = 0$, multiplied by $(M_S/\gamma)$ to make it dimensionless, as

$$\lambda_E^* = \frac{h_S}{\cos\theta \sin\theta - h_x \cos\theta} - 1. \quad (8)$$

Then, the thermal energy barrier $E_B$ is obtained by [42]

$$E_B = -V \int_{E_{\min}}^{E_{\max}} dE \, \lambda_E^*, \quad (9)$$

where $E_{\max}$ and $E_{\min}$ are obtained from $\lambda_E^* = 0$.

Figure 2(a) shows $\lambda_E^*$ versus $E$ for $h_s = h_x = 0.2$, where $E_{\max}$ and $E_{\min}$ are identified. The integral of Eq. (9) in the region $\lambda_E^* < 0$ gives the energy barrier. Figure 2(b) shows $\lambda_E^*$ versus $E$ for $h_x = 0.2$ and various values of $h_s$. We note that both $E_{\max}$ and $E_{\min}$ vary with $h_x$ and $h_s$. In Fig. 2(b), the area for $\lambda_E^* < 0$ becomes zero at $h_s = 0.364$. For these parameters, therefore, the energy barrier $E_B$ vanishes at the current corresponding to $h_s = 0.364$. The threshold current corresponding to this threshold $h_s$ vanishing $E_B$ should be equivalent to the threshold current described by Eq. (1). We test this correlation and find that they are indeed equivalent (see the inset of Fig. 3(a)).

An analytic expression relating $E_B$ to $h_s$ is essential to design and interpret experimental results, e.g. switching current as a function of the current pulse-width. By solving Eq. (8) = 0 and Eq. (9), one can find the analytic expression of $E_B$, given as

$$E_B = E_B^0 F(h_x, h_s), \quad (10)$$

where $E_B^0 = H_{K,\text{eff}} M_S V / 2$. The factor $F(h_x, h_s)$ describes how the energy barrier is influenced by the external magnetic field ($h_x$) and current ($h_s$). The general form of $F(h_x, h_s)$ can be derived, but it is too complicated to be useful practically. We derive an approximated analytic expression of $F(h_x, h_s)$ up to the 2nd order of ($h_x$, $h_s$) assuming ($h_x$, $h_s$) << 1. It is given as

$$F(h_x, h_s) = \left[(1-h_x)^2 - 2h_s\left(\frac{\pi}{2} - h_x - h_s\right)\right]. \quad (11)$$

Figure 3(a) shows that exact solutions (symbols) and approximated ones (lines; Eq. (11)) of $F(h_x, h_s)$ are in reasonable agreement for wide ranges of ($h_x$, $h_s$). The approximated expression of $F(h_x, h_s)$ has a quasi-linear dependence on $h_s$ (thus current), since $h_s$ << $\pi/2$. It is instructive to note that the $E_B$ of SO-STT switching described in Eqs. (10) and (11) is clearly different from the $E_B$ of the standard STT switching given as $E_B^0 (1 - I/I_c)^2$ [43-45] where $I$ and $I_c$ are the applied current and the threshold current at $T = 0$ K, respectively.

The expression of energy barrier, $E_B^0 F(h_x, h_s)$, provides an analytic form of threshold current for thermally activated switching of a perpendicular magnet by SO-STT. The threshold $h_s^{\text{th}}$ for thermally activated switching by SO-STT can be obtained from Eqs. (10)-

(11) and the switching probability $P_{SW}$ of 1/2, where $P_{SW} = 1 - \exp[-f_0 t \exp(-E_B/k_B T)]$, $f_0$ is the attempt frequency, and $t$ is the measurement time (i.e., current pulse-width);

$$h_s^{th} = \frac{1}{4}\left(\pi - 2h_x - \sqrt{8\left(\frac{\ln(f_0 t/\ln(2))}{\Delta} - 1\right) - 4h_x^2 - 4h_x(\pi - 4) + \pi^2}\right), \quad (12)$$

where $\Delta = E_B^0/k_B T$. Using $h_S = c_J/H_{K,eff}$ and $c_J = (\hbar/2e)(\theta_{SH}^{eff} J/M_S t_F)$, one finds the threshold current $I_{SW}^{SO-STT}$ for thermally activated switching due to SO-STT as

$$I_{SW}^{SO-STT} = \frac{eH_{K,eff} M_S t_F A_{HM}}{2\hbar \theta_{SH}^{eff}}\left(\pi - 2h_x - \sqrt{8\left(\frac{\ln(f_0 t/\ln(2))}{\Delta} - 1\right) - 4h_x^2 - 4h_x(\pi - 4) + \pi^2}\right). \quad (13)$$

Equation (13) is the key result of our work. Figure 3(b) shows $h_s^{th}$ as a function of the log-scaled current pulse-width ($\log(t)$) for $\Delta = 50$ and $f_0 = 1$ GHz with varying $h_x$. It is noteworthy that $h_s^{th}$ is in almost linear relation to $\log(t)$ even below 10 ns. This linear relation of the SO-STT switching even at short pulse-width is in sharp contrast to the standard STT switching in which the switching current rapidly increases at short pulses due to the precessional switching mode.

To stress this difference between the SO-STT switching and standard STT switching at short pulses, we calculate the switching current as a function of pulse-width (Fig. 4(a)) using macrospin simulations with including the thermal fluctuation fields, i.e., the Gaussian-distributed random fluctuation fields (mean = 0, standard deviation = $\sqrt{2\alpha k_B T/(\gamma M_S V \delta t)}$, where $\delta t$ is the integration time step [46]). In this macrospin simulation, we use the following parameters; the gyromagnetic ratio $\gamma = 1.76 \times 10^7$ Oe$^{-1}$s$^{-1}$, the Gilbert damping constant $\alpha = 0.1$, the saturation magnetization $M_S = 1000$ emu/cm$^3$, the free layer thickness $t_F = 2$ nm, the area of nanopillar is $\pi(15$ nm$)^2$, and the rise/fall time of the current pulse is 0.5 ns. For the standard STT switching, we assume the spin polarization factor $P = 0.7$ and the effective perpendicular anisotropy field $H_{K,eff}$ = 2344 Oe. For the SO-STT switching, we assume the effective spin Hall angle $\theta_{SH}^{eff}$ = 0.3, the cross-sectional area of heavy metal layer $A_{HM}$ = (30 nm × 2 nm), $H_{K,eff}$ = 4100 Oe, and $H_x$ = 1000 Oe. These choices of $H_{K,eff}$ and $H_x$ give the same energy barrier $\Delta$ = 40 at zero current for both standard STT switching and SO-STT switching.

Figure 4(a) summarizes the simulation results. Each data point is obtained from $P_{SW}$ = 1/2 at a given current pulse-width with more than 2,000 switching trials. At long pulses, the switching current $I_{SW}$ of the standard STT switching is smaller than the $I_{SW}$ of the SO-STT switching. As well known, the $I_{SW}$ of the standard STT switching rapidly increases as the current pulse becomes shorter than 10 ns. By contrast, the $I_{SW}$ of the SO-STT switching does not exhibit such a rapid increase at short pulses. This quasi-linear change in the $I_{SW}$ of the SO-STT switching even at short pulses is reasonably well described by Eq. (13) (i.e., the blue solid line in Fig. 4(a); here we assume $f_0$ = 0.7 GHz, a single unknown parameter in Eq. (13)). There are small deviations between the modeling results and ones predicted from Eq. (13) at the current pulse-width below 10 ns, which we attribute to a finite rise/fall time in simulations. The results shown in Fig. 4(a) demonstrate that (i) Eq. (13) captures the core effects of thermally activated perpendicular magnetization switching due to SO-STT, and (ii) the SO-STT switching can yield a smaller switching current than the standard STT switching at short pulses, which is beneficial for fast MRAM operations.

Based on Eq. (13), we next estimate the switching probability of a fully-selected cell ($P_{SW}^{Full}$) and a half-selected cell ($P_{SW}^{Half}$) of the SO-STT switching. For the memory applications with SO-STT, the two switching probability curves, i.e., $P_{SW}^{Full}$ and $P_{SW}^{Half}$ as a function of the current, must be sufficiently separated along the current axis. We assume that a voltage is applied to the fully selected cell, resulting in the voltage-induced anisotropy reduction of $\Delta H_{\text{K,eff}}$. We test three cases of $\Delta H_{\text{K,eff}}$ (= 0.5, 1, and 2 kOe; see Fig. 4(b)) at the current pulse-width of 10 ns. For other parameters, we assume the same values used for Fig. 4(a). For $\Delta H_{\text{K,eff}}$ = 0.5 kOe, the $P_{\text{SW}}^{\text{Full}}$ and $P_{\text{SW}}^{\text{Half}}$ curves are not sufficiently separated. For $\Delta H_{\text{K,eff}}$ = 2 kOe, on the other hand, the separation between the $P_{\text{SW}}^{\text{Full}}$ and $P_{\text{SW}}^{\text{Half}}$ curves are large enough to ensure a wide window for the writing current. In Ref. [25], we showed that $\Delta H_{\text{K,eff}}$ of about 3 kOe is required to reduce the switching current density below $10^7$ A/cm$^2$ when the voltage-induced anisotropy change is accompanied by SO-STT as a write scheme of MRAMs. The results in Fig. 4(b) suggest that such enhanced voltage-induced anisotropy change is also essential to secure a wide writing current window.

To summarize, we obtain the analytic expression of threshold current for thermally activated switching of a perpendicular magnet by SO-STT. The expression is valid even below 10 ns, because of non-precessional SO-STT switching. Based on the expression, we find that the thermal energy barrier shows a quasi-linear dependence on SO-STT, and consequently, the threshold switching current depends on the log-scaled current pulse-width almost linearly. Our results will be useful to design memory devices utilizing SO-STT and to interpret SO-STT switching experiments with nanomagnets.

This work was supported by the NRF (2011-0028163, NRF-2013R1A2A2A01013188), and Human Resources Development Program, MKE/KETEP (No. 20114010100640).

**FIGURE CAPTION**

FIG. 1. (Color online) A schematic architecture of perpendicular SO-STT MRAM consisting of multiple MRAM cells sharing the same heavy metal line and select-transistors (or diodes) to apply a voltage to individual cells.

FIG. 2. (Color online) (a) Dependence of $\lambda_E^*$ on the energy $E$ normalized by $M_S H_{K,\text{eff}}/2$ for the SO-STT switching of a perpendicular magnet for $h_s = h_x = 0.2$. (b) Dependence of $\lambda_E^*$ on the energy $E$ normalized by $M_S H_{K,\text{eff}}/2$ for $h_x = 0.2$ with varying $h_s$.

FIG. 3. (Color online) (a) $F(h_x, h_s)$ as a function of $h_s$ and various values of $h_x$. Symbols are exact solutions obtained from Eq. (8) = 0 and Eq. (9). Lines are obtained from Eq. (11). Exact solutions and Eq. (11) are in reasonable agreement for wide ranges of $h_x$ and $h_s$. (b) The threshold $h_s^{\text{th}}$ versus log-scaled $t$ for $\Delta = 50$, $f_0 = 1$ GHz, with varying $h_x$, obtained from Eq. (13). Inset of (a) shows threshold $h_s$ as a function of $h_x$. Symbols are obtained from Eqs. (8) and (9), for the condition of vanishing $E_B$. Solid line is obtained from Eq. (1). The two results match exactly.

FIG. 4. (Color online) (a) Switching current $I_{SW}$ as a function of the current pulse-width. Symbols are obtained from macrospin simulations. Blue solid line is obtained from Eq. (13) with assuming $f_0 = 0.7$ GHz. (b) Switching probability curves of a fully selected cell ($P_{SW}^{\text{Full}}$) and a half-selected cell ($P_{SW}^{\text{Half}}$), obtained from Eq. (13). For $P_{SW}^{\text{Full}}$, three cases with different voltage-induced anisotropy change $\Delta H_{K,\text{eff}}$ (= 0.5, 1, and 2 kOe) are considered.

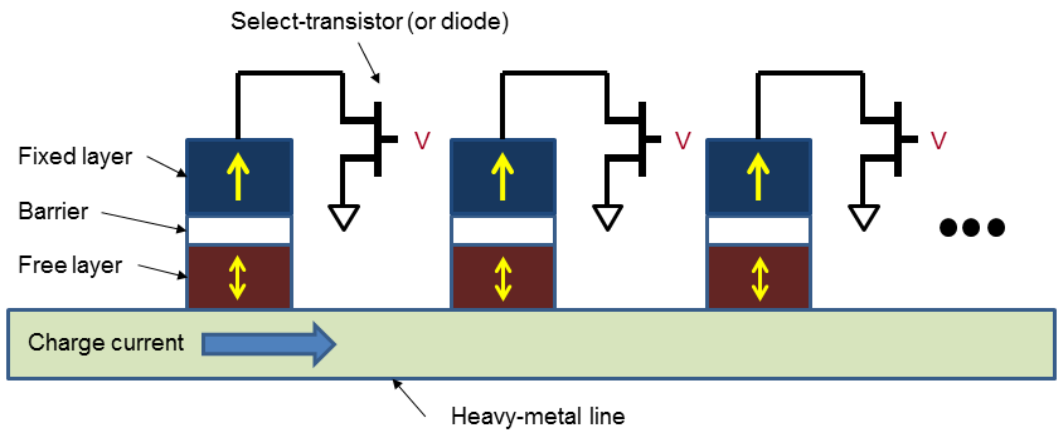

FIG. 1. Lee et al.

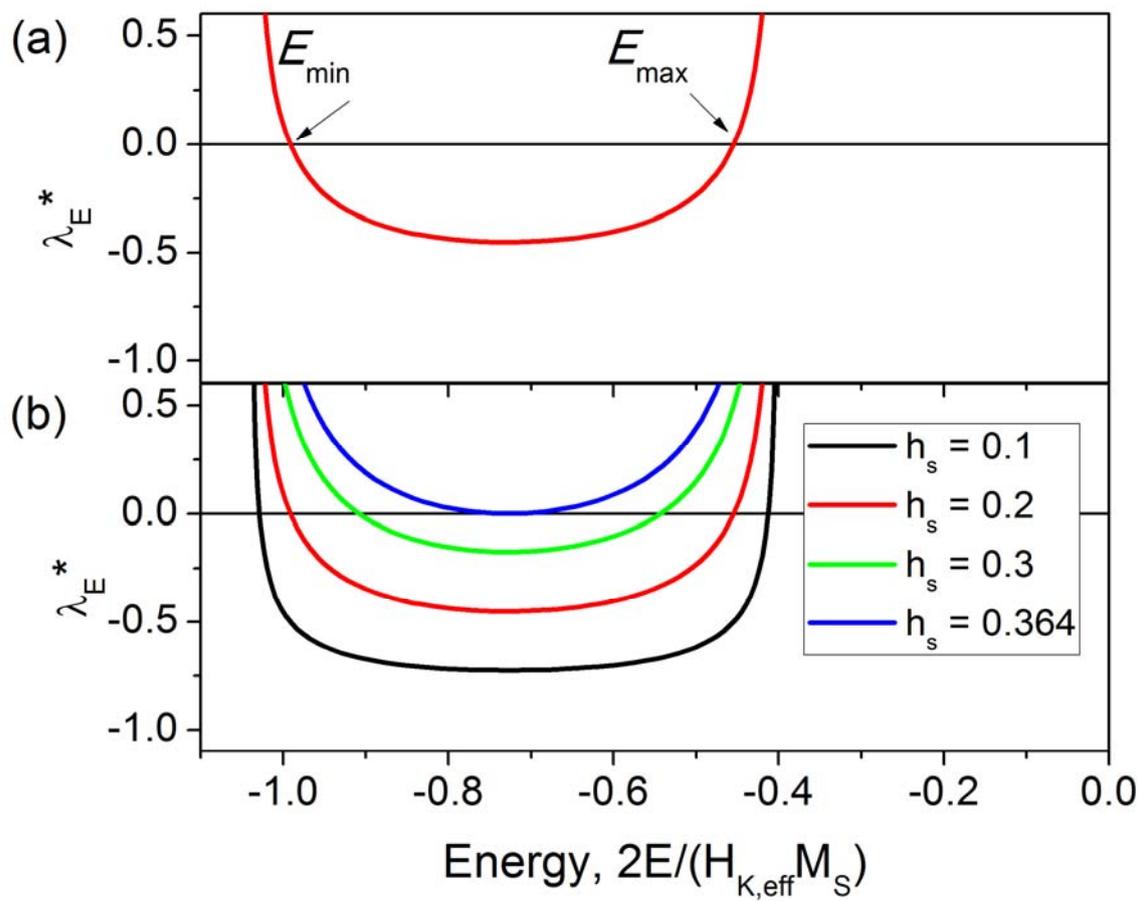

FIG. 2. Lee *et al.*

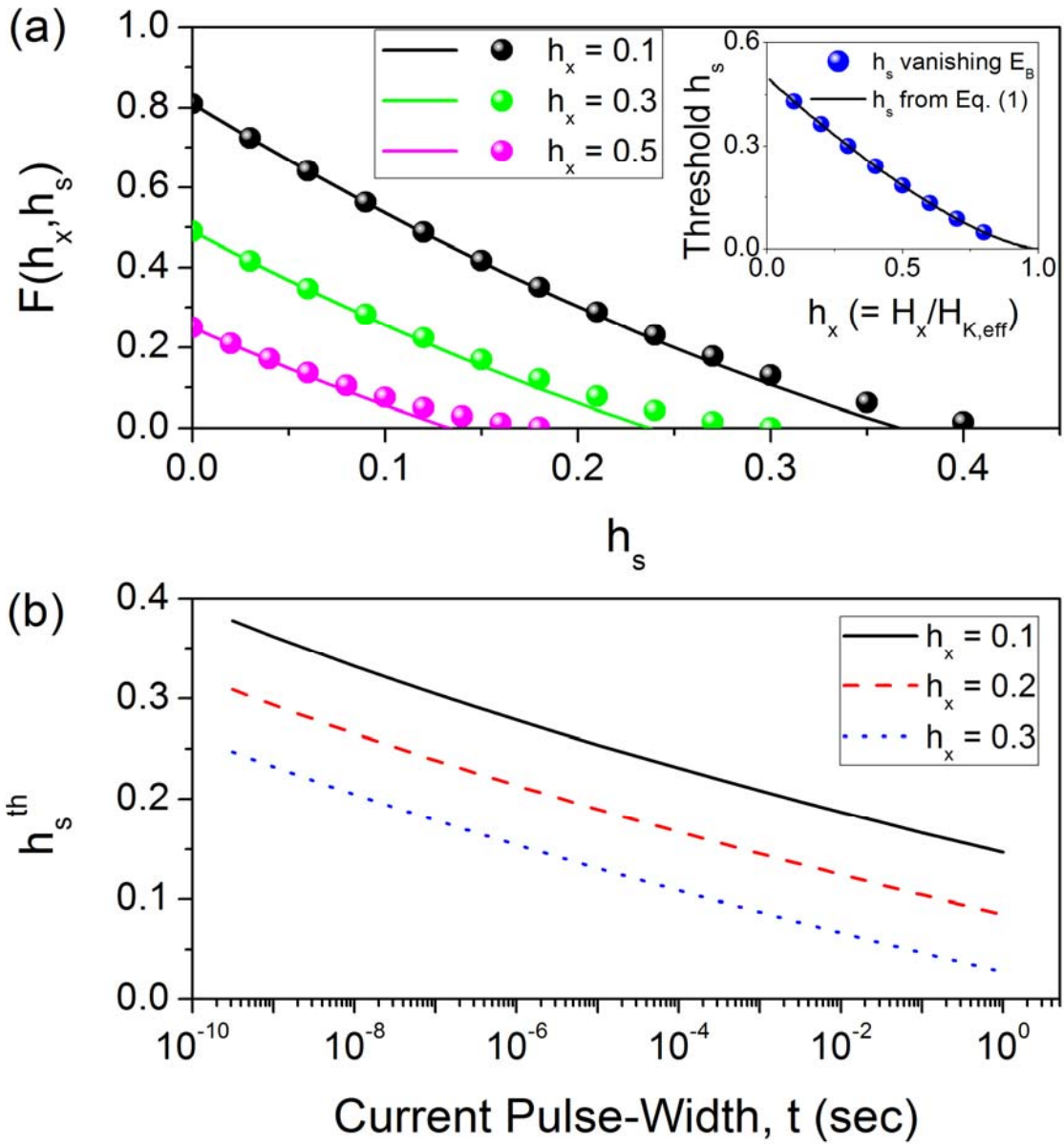

FIG. 3. Lee *et al*.

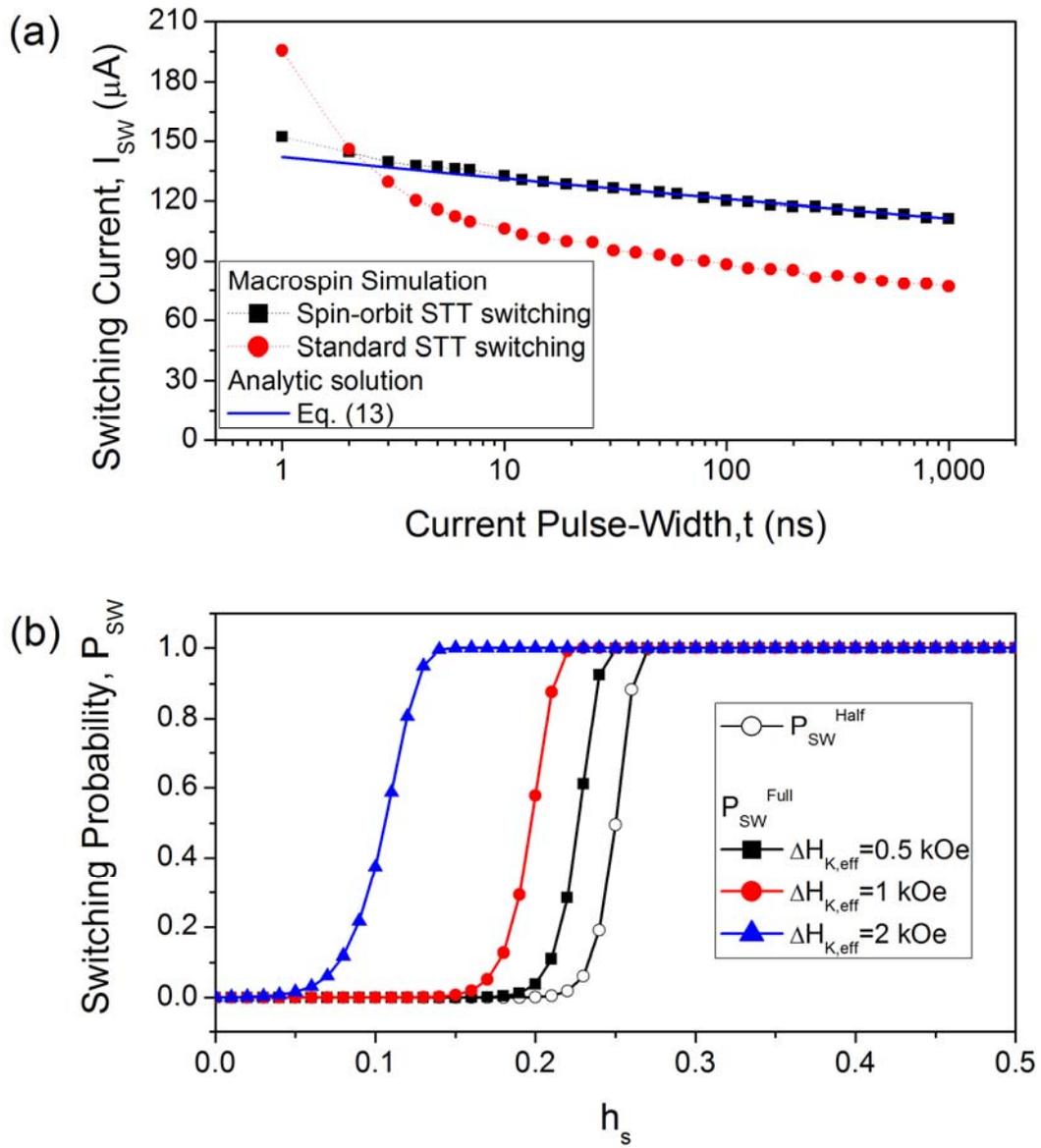

FIG. 4. Lee *et al*.